

\input phyzzx.tex
\def\a{\alpha}
\def\b{\beta}

\def\d{\delta}

\def\z{\zeta}

\def\r{\rho}

\def\k{\kappa}
\def\l{\lambda}

\def\o{\omega}

\def\s{\sigma}

\def\pa{\partial}

\def\ov{\overline}

\def\zp{\zeta_+}
\def\zm{\zeta_-}

%
\def\ap#1{{\it Ann. Phys.} {\bf #1}}
\def\cmp#1{{\it Comm. Math. Phys.} {\bf #1}}

\def\pl#1{{\it Phys. Lett.} {\bf B#1}}

\def\prd#1{{\it Phys. Rev.} {\bf D#1}}

\def\np#1{{\it Nucl. Phys.} {\bf B#1}}

%

\REF\call{C.G. Callan, S.B. Giddings, J.A. Harvey, and A. Strominger,
\prd{45} (1992) R1005.}
\REF\sda{S. P. de Alwis, \pl{289} (1992) 278; \pl{300} (1993) 330 .}
\REF\bc{A. Bilal and C. Callan, \np{394} (1993) 73.}
\REF\shda{S. P. de Alwis, \prd{46} (1992) 5429.}
\REF\rst{J.G. Russo, L. Susskind, L. Thorlacius, \pl{292} (1992) 13;
J. Russo, L. Susskind, and L. Thorlacius, \prd{46} (1992) 3444.}
\REF\hawk{S. W. Hawking, \cmp{43} (1975) 199.}
\REF\mc{S. P. de Alwis and D. A. MacIntire, \prd{50} (1994) 5164.}
\REF\banks{V. Lapchinsky and V. Rubakov, {\it Acta Phys. Pol.} {\bf B
10}
(1979) 1041;\hfill\break T. Banks, \np{249} (1985) 332.}
\REF\sam{E. Keski-Vakkuri, G. Lifschytz, S. D. Mathur, M. E. Ortiz,
preprint
MIT-CTP-2341, hep-th/9408039 (1994).}
\REF\dea{S. P. de Alwis, \pl{317} (1993) 46. }
\REF\sv{K. Schoutens, E. Verlinde and H. Verlinde, \prd{48} (1993)
2690.}
\REF\strom{A. Strominger, \prd{48} (1993) 5769.}
\REF\fiola{T. Fiola, J. Preskill, A. Strominger, S. Trivedi, \prd{50}
(1994)
3987.}
\REF\binir{B. Binir, S. B. Giddings, J. A. Harvey, A. Strominger,
\prd{46}
(1992) 638.}
\REF\rt{T. Regge and C. Teitelboim, \ap {88},  286 (1974).}
\REF\dijkvv{R. Dijkgraaf, H. Verlinde, E. Verlinde, \np{371} (1992)
269; S.
Hirano, Y. Kazama, Y. Satoh, \prd{48} (1993) 1687.}

\pubnum {COLO-HEP-341\cr hep-th/9411003}
\date={October 1994}
\titlepage
\vglue .2in
\centerline{\bf Lessons of Quantum 2D Dilaton Gravity }
\centerline{\bf  }
\author{ S.P. de Alwis\foot{dealwis@gopika.colorado.edu} and D.A.
MacIntire\foot{doug@dam.colorado.edu}}
\address{Dept. of Physics, Box 390,\break
University of Colorado,\break Boulder, CO
80309}
\vglue .2in
\centerline{\caps ABSTRACT}
Based on the analysis of   two dimensional dilaton gravity  we argue
that the
semiclassical equations of black hole formation and evaporation
should not be
interpreted in terms of   expectation values of operators in the
exact  quantum
theory, but rather as WKB trajectories. Thus at the  semiclassical
level it
does not seem possible to formulate a notion of {\it quantum
mechanical}
information loss.
\endpage

The development of two-dimensional dilaton gravity models (CGHS)
[\call\ -
\rst]
raises the possibility of understanding some conceptual issues in
quantum
gravity. In particular it may be hoped that the question of
information loss in
the evaporation of black holes [\hawk ] might be resolved within this
simple
model. Another issue that may be addressed is  the problem of time in
a quantum
theory of gravity.  Actually, as discussed in [\mc], in order to
resolve the
first issue it is necessary to have some understanding of the second.

In the literature on quantum gravity the semiclassical approximation
is often
confused with the Born-Oppenheimer approximation. As shown in [\banks
] the
latter approximation (which for 4D gravity consists of expanding in
powers of
$M_P^{-1}$), when applied to the Wheeler-DeWitt (WDW) equation, leads
to a
Schr\"odinger equation for the matter wave function in a background
geometry
that is a solution of the vacuum Einstein equation.
However, in much of the work on black hole evaporation that attempts
to include
the effect of backreaction, it is assumed that the gravitational
field may be
treated classically according to the equation ${G_{\mu\nu}=8\pi
G<T_{\mu\nu}>}$,  where the expectation value is to be taken in a
quantum state
of matter that satisfies the Schr\"odinger equation.  Although many
authors
have attempted to obtain this from the WDW equation,  there is no
real
derivation of this approximation. In fact the source of the problem
in our
opinion is the appearance of an expectation value, i.e. a quantity
that
violates the superposition principle. \foot {The well known argument
for
information loss assumes that there is a notion of time evolution of
quantum
states defined on a series of space-like surfaces in an evaporating
black hole
space time that is a solution of the semiclassical equation. It is
then
implicitly assumed (in the case of the argument for no information
doubling the
assumption is explicit) that the superposition principle is valid for
the
matter quantum state. This is however not compatible with the
assumption of a
definite classical solution determined by the {\it expectation value}
of the
stress tensor. For a concrete calculation illustrating the problems
involved
see [\sam]. } Clearly, starting from a linear equation like the WDW
equation,
it is unlikely that there can be a satisfactory derivation of a
non-linear
``approximation''.\foot{For a recent review of the problems involved
see
[\mc].}

How then are we to understand the equations that have been discussed
in two
dimensional dilaton gravity?  What is the interpretation of these
so-called
semiclassical equations?  How can they be obtained from an exact
formulation of
the theory?  In [\dea, \mc ] the problems associated with what is
usually
called the semiclassical approximation were discussed\foot{Actually
in that
paper the discussion at the end was couched in terms of
deBroglie-Bohm
trajectories, which of course are supposed to describe the complete
quantum mechanical situation, and therefore involved potentially
controversial
interpretational issues.  However, for the purpose of just discussing
the
emergence of the
semiclassical physics, it is sufficient to consider the leading order
approximation, which  just gives the well-known WKB trajectories.}.
In this
note we discuss other possibilities for deriving a semiclassical
interpretation, including the use of coherent states that satisfy the
constraints.  We argue that the only consistent interpretation of the
semiclassical equations found in the literature is to regard these
equations as
defining the WKB trajectories of the Wheeler-DeWitt wave function.
We also
show that the quantum analog of the ADM Hamiltonian (that may be
taken to be
conjugate to the time measured by clocks at infinity) in these
dilaton gravity
theories plays no role in the local physics.   Finally we discuss the
implications for the problem of information loss.

In conformal gauge, the Liouville-like version of  the CGHS action,
which
incorporates corrections coming from the functional integral measure,
is given
by [\sda, \bc ]
$$S={1\over 4\pi}\int d^2\s [\mp\pa_{+}X\pa_{-}
X\pm\pa_{+}Y\pa_{-}Y+\sum_{i=1}^N
\pa_{+}f_i\pa_{-}f_i+2\l^2e^{\mp\sqrt{2\over \k} (X\mp Y)}].
\eqn\action$$
Here  $\k ={N-24\over 6}$, $N$ being the number of matter
fields\foot{ Note
that this definition has the opposite sign to the $\k $ defined in
[\sda].}.
We have omitted in the above the ghost action that comes from gauge
fixing to
the conformal gauge since we are
going to discuss only semiclassical, i.e. large N, effects. Thus for
the
purposes of this paper $\k $  may be replaced by $N/6 $.  Also in the
above we
will choose the lower sign corresponding to $\k > 0$.   The field
variables in
the above are related to the original variables $\phi$ and $\rho$
that occur in
the CGHS action, gauge fixed to the conformal gauge
$(g_{\a\b}=e^{2\rho}\eta_{\a\b})$, through the following relations:
$$Y=\sqrt{2\k}\left[\r+\k^{-1}e^{-2\phi}-{2\over\k}\int d\phi
e^{-2\phi}\ov h
(\phi)\right],
\qquad X=2\sqrt{2\over \k}\int d\phi P(\phi ),\eqn\xycdt$$
where $P(\phi )=e^{-2\phi}[(1+\ov h)^2-\k e^{2\phi}(1+h)]^{1\over
2}$.
In  \xycdt, the functions $h(\phi),~\ov h(\phi)$ parametrize quantum
(measure)
corrections that may come in when defining the theory with respect to
a
translationally invariant measure (see [\shda]  for details).

The discussion of Hawking evaporation has been based on the solution
of the
classical equations of motion coming from \action\ and imposing the
classical
constraint equations:
${T^{X,Y}_{\pm}+T^f_{\pm}=0}$.   What interpretation do these
equations have
from the standpoint of the exact quantum theory of \action?  It has
already
been pointed out [\dea, \mc ] that these equations cannot be
interpreted as the
expectation values of the field operators in physical
states.\foot{The physical
states are normally defined as solutions of the Wheeler-DeWitt
equation or the
BRST condition $Q_{BRST}\Psi =0 $.} The reason is the well-known
problem of
time in quantum gravity.  How then are these equations to be
interpreted.

The first issue is whether it is possible to derive the semiclassical
equations
starting from (say) coherent states of the DDF operators constructed
in [\sv,
\dea ].  The DDF operators ($\hat A$) are gravitationally dressed
creation and
annihilation operators
for the $f$ fields that commute with the total stress-tensor
operator. Thus
states constructed from them (by operating on the Fock vacuum) will
satisfy the
physical state
condition (at least in the form $T^{+}_{total}|\Psi>=0$ or the BRST
form, if
not in the Wheeler-DeWitt form) and one may ask whether the
semiclassical
equations have an interpretation as expectation values in these
states. Now as
pointed out in [\dea ], the expectation values of the field operators
$\hat f$,
$\hat X$, $\hat Y$ in these states are independent of time  if the
time
translation generator is the usual one, namely the spatial integral
of the
time-time component of the total stress tensor. However one might use
a
different Hamiltonian as the time translation generator. In fact, one
might
think that the natural Hamiltonian to use is $\hat H_d=\int d\o\,\o
\hat
A^{\dag}(\o )\hat A(\o )$. Clearly the field operator
$$f_d(\s^+,\s^-)=\int_{\o >0}d\o\left[A_+(\o
)e^{-i\o\s^+}+A_+^{\dag}(\o
)e^{i\o\s^+} + \; ({\scriptstyle + \; \rightarrow \;
-})\right],\eqn\fdress$$
which satisfies the free field equation $\pa_+\pa_-f_d =0$, evolves
in time
according to the Heisenberg equation $\dot f_d=i\,[H_d,f_d]$.
Coherent states
may then be defined as (using a somewhat schematic notation)
$$|f_c>=:e^{\int d\s f_c\pa f_d}:|0>,\eqn\coh $$
and the classical field $f_c $ is the expectation value of the
dressed field
operator in this state. However, it is not possible to obtain the
rest of the
semiclassical system of equations. In particular the time evolution
of the
geometry (i.e the fields $X$ and $Y$) as defined by this new
Hamiltonian is now
given by a non-local equation since the Hamiltonian density is
non-local in
those fields\foot{This is easily seen by examining the definitions
for the DDF
operators, see Refs.  [\sv, \dea ].} .  Also, even though the
constraint
equation $T_{tot}^{(+)}|\Psi>=0$ implies that
$<f_c|T^{X,Y}+T^{f}|f_c>=0$, the
usual semiclassical equations cannot be obtained from this since that
would
mean again taking the expectation values of $f$, $X$, $Y$, etc. in
the physical
states, and  then we are back to the old problem of the time
independence of
the expectation values.

In some recent papers (for instance [\strom, \fiola ]) the equations
of the
theory are interpreted in the following fashion. Consider the
constraint
equation in the Kruskal gauge (which in our notation means the choice
of
coordinates such that $X+Y=0$): $$\sqrt{\k\over
2}\pa^2_{\pm}X(x^{\pm})=\pa_{\pm}f_c\pa_{\pm}f_c.$$
It is then proposed that the left hand side of this equation (i.e.
the
geometrical fields) be treated as classical fields and the right hand
side as
the expectation value of the matter field operator in a coherent
state of the
$f$-field\foot{I.e. \coh\ with $\hat f_d$ replaced by $\hat f $}. But
as
discussed in detail in [\mc ] this is precisely the interpretation
that cannot
be obtained if one starts from the Wheeler-DeWitt equation (or any
other
implementation of the  constraint as a linear condition on the
states). An
alternative that one might consider is the imposition of a non-linear
condition
$<T^{total}>=0$ as was proposed in [\dea]. Here one has a
Schr\"odinger
evolution of the states, but the linear superposition principle has
been
abandoned. It is not clear whether this is a viable theory of quantum
gravity,
but it is perhaps the only option available if one wants to interpret
the
semiclassical theory in terms of expectation values of field
operators.
However there are still ambiguities in the treatment of the
constraints that we
wish to highlight now.

The states we wish to consider are coherent states of the matter and
dilaton
gravity fields. These may be built either on the vacuum that is
annihilated by
operators defined with respect to the Kruskal coordinates $x^{\pm}$,
or by that
defined with respect to the sigma coordinates
$\s^{\pm}=\pm\l^{-1}\ln{(\pm\l
x^{\pm})}$.  Thus we have two possible states to consider: $|,,,K>$
or
$|,,,\s>$, where the commas denote the values of the classical field
configurations for the $X$, $Y$, and $f$ fields. However there are
two possible
operators to consider as well.  Namely
$$\hat T^{X,Y}+:\hat T^f:_K,\eqn\kop$$and
$$\hat T^{X,Y}+:\hat T^f:_{\s}.\eqn\sop$$
Note that these operators are just  the expressions for the total
stress tensor
operator, apart from the ghost stress tensor that we  ignore since we
are just
interested in the large $N$ limit.  In this same limit we may ignore
the normal
ordering of the $X,Y$ stress tensor.  Now if we are to consider the
physical
state conditions as the equation that the expectation values of these
operators
is zero, then clearly there are four possibilities.  Equating the
expectation
value in the Kruskal states of the Kruskal normal-ordered operator
\kop\ to
zero we have,
 $$-\sqrt{\k\over 2}\pa_{\pm}^2\ov X(x) + \pa_{\pm}\ov f\pa_{\pm}\ov
f(x) =
0.\eqn\stat$$
The fields in the above equation are the classical fields out of
which the
coherent states are constructed. The fields in the Kruskal gauge are
barred to
distinguish their functional form from that of the fields in the
sigma gauge.
Of course, since all the fields in the above equation are scalars, we
have $\ov
X(x)=X(\s)$  and $\ov f(x) =f(\s )$.  Similarly, equating the
expectation value
in the sigma states of the sigma normal ordered operator \sop, we
have
$$-\left[\sqrt{\k\over 2}(\pa_{\pm}^2 X\mp\pa_{\pm}X)(\s )
+{\k\l^2\over
4}\right]+\pa_{\pm} f\pa_{\pm}f(\s )=0.\eqn\rad$$
If one now uses also the equations of motion for these fields then it
is
possible to show that the relation \rad\ is the one  corresponding to
Hawking
radiation [\call - \rst ] whilst \stat\ seems to describe the static
situation
of a black hole in equilibrium with a radiation bath [\binir ].

On the other hand one may also evaluate the Kruskal normal-ordered
operator
\kop\ in the sigma states using
$$:\hat T^f:_{\s}(\s )=\left ({dx\over d\s}\right )^2:\hat
T^f:_K(x)-{\k\l^2\over 4}$$
and get
$$-\sqrt{\k\over 2}\left(\pa_{\pm}^2 X\mp\pa_{\pm}X\right)(\s )
+\pa_{\pm}
f\pa_{\pm}f(\s )=0. \eqn\static$$
This is nothing but the coordinate transformed version of \stat\ and
describes
the static situation. Similarly, evaluating  the sigma normal ordered
operator
\sop\ in the Kruskal states gives us
$$-\sqrt{\k\over 2}\pa_{\pm}^2\ov X(x) + \pa_{\pm}\ov f\pa_{\pm}\ov
f(x) =
0,\eqn\radn$$
which is of course just \rad\ evaluated in Kruskal coordinates.

The moral of the story is that it is irrelevant which state the
expectation
value is taken, only the operator appears to be  relevant. This
should be
compared with the usual notion,
obtained from intuition derived from working in a fixed background,
that the
two vacua (the Hartle-Hawking one, i.e. the one that is annihilated
by the
positive  frequency modes defined in the Kruskal gauge, and the Unruh
vacuum
defined relative to the sigma coordinates) are what determine the two
different
physical situations. In any case, the fact that one has to abandon
the
superposition principle in order to justify this expectation  value
interpretation makes it unattractive. Also one cannot within this
interpretation formulate the usual arguments for information loss
which depend
on being able to deal with (a  linear subspace of ) the  Hilbert
space at least
for the matter sector.

This brings us to the  interpretation of the semiclassical equations
as WKB
trajectories.
To facilitate the (formal) justification of our approximations it is
convenient
to explicitly introduce $\hbar$ into the discussion. Thus in all of
the above
equations $\k$ should actually be replaced by $\k\hbar $: these terms
come
from measure corrections and thus have an explicit factor of $\hbar$.
Now the
$1\over\k$ expansion is done keeping $\k\hbar $ fixed at $O(1)$, and
therefore
it is nothing but the usual (WKB) semiclassical expansion in powers
of $\hbar
$.  In fact it is convenient to choose $\k\hbar =1 $. Introducing the
fields
$\zeta_+ = \ln{2} + \sqrt{2}\ (X+Y)$ and $\zeta_- = \sqrt{2}\ (X-Y)$,
the
(total) stress tensors then take the form (ignoring the ghosts)
$$\eqalign{T_{\pm\pm} &= {1\over 4}\left[ \partial_\pm \zeta_+
\partial_\pm
\zeta_- +
\partial_\pm^2 (\zeta_+ - \zeta_-)\right] + {1\over 2}\sum_{i=1}^N
\partial_\pm
f_i \partial_\pm f_i\cr
T_{+-} &= -{1\over 4} \partial_+\partial_-  (\zeta_+ - \zeta_-)  -
{1\over2}\lambda^2 e^{\zeta_+ }.\cr}$$
Using the canonical momenta $\Pi_\pm = {1\over 16\pi} \dot
\zeta_\mp$,
$\Pi_{f_i}= {1\over 8 \pi} \dot f_i $, the Hamiltonian and momentum
densities
are:
$$\eqalign{T_{00} &= {32\pi^2} \Pi_+ \Pi_- + {1\over 8}\left[\zeta'_+
\zeta'_-
+ 2(\zeta''_+  - \zeta''_-) - 8\lambda^2 e^{\zeta_+}\right] + {1\over
4}\sum_{i=1}^N\left[ {f'_i}^2 + 16\pi^2 \Pi^2_{f_i}\right]\cr
T_{01} &=2\pi\left[\zm' \Pi_-+ \zp'\Pi_+  + 2{\pa\over \pa \s}
\left(\Pi_- -
\Pi_+ \right)+ \sum_{i=1}^N f'_i \Pi_{f_i}\right].}$$

We now quantize using $\Pi_u \rightarrow {\hbar\over i} {\delta\over
\delta u}$
to obtain the corresponding operators $\hat T_{00}$ and $\hat
T_{01}$.  The
quantum mechanical constraint (Wheeler-DeWitt) equations become:
$$ \hat T_{00} \Psi\left[\zeta_+,\zeta_-,f_i\right] = 0,\qquad
{\rm and}\qquad \hat T_{01} \Psi\left[\zeta_+,\zeta_-,f_i\right] =
0.\eqn\wdwdgdiffeo$$
Writing $\Psi$ in the  form
$$\Psi = R\, [\zeta_+,\zeta_-,f_i] \exp{\left\{{i\over\hbar
}S\,[\zeta_+,\zeta_-,f_i] \right\}}$$
we obtain for the real part of $\hat T_{00} \Psi =0$,
$${32\pi^2} {\d S \over \d\zp} {\d S \over \d\zm}  +
V[\zeta_+,\zeta_-] +
V_m[f']  +
Q +  4 \pi^2 \sum_{i=1}^N \left({\d S \over \d f_i}\right)^2 =
0\eqn\ReTzz$$with
$$V[\zeta_+,\zeta_-] = {1\over 8} \left[\zeta'_+\zeta'_- +
2(\zeta''_+  -
\zeta''_-) - 8\lambda^2 e^{\zeta_+}\right],\qquad V_m[f']={1\over
4}\sum_{i=1}^N {f'_i}^2, $$
$$Q =-\hbar^2\left [ {32\pi^2\over \k} {1\over R} {\d^2 R\over \d\zp
\d\zm}  +
{4\pi^2\over R}\sum_{i=1}^N {\d^2 R\over \d f_i^2}\right ] .$$
The imaginary part of  $\hat T_{00} \Psi =0$ gives
$$ \sum_{i=1}^N {\d \over \d f_i} \left(R^2 {\d S\over \d f_i}\right)
+ {4}
\left[  {\d \over \d \zp}\left(R^2 {\d S\over \d \zm}\right) + {\d
\over \d
\zm}\left(R^2 {\d S\over \d \zp}\right)\right] = 0.\eqn\ImTzz$$

Now let us write out the WKB (semiclassical) expansion  as,
$$\eqalign{S[\zeta_+,\zeta_-,f_i] &= S_{-1}[\zeta_+,\zeta_-,f_i] +
\hbar
S_0[\zeta_+,\zeta_-,f_i]  + \hbar^2 S_1[\zeta_+,\zeta_-,f_i] +
\ldots\cr
R[\zeta_+,\zeta_-,f_i] &= R_0[\zeta_+,\zeta_-,f_i]  +
{\hbar}R_1[\zeta_+,\zeta_-,f_i] + \ldots.\cr}$$
Then the leading order WKB approximation to the Wheeler-DeWitt wave
function is
$$\Psi\simeq\Psi_{WKB }=R_0e^{{i\over\hbar}S_{-1 }},$$
where
$${32\pi^2} {\d S_{ -1} \over \d\zp} {\d S_{ -1} \over \d\zm}  +
V[\zeta_+,\zeta_-] + V_m[f']  +  4 \pi^2 \sum_{i=1}^N \left({\d S_{
-1}  \over
\d f_i}\right)^2 = 0,\eqn\hamjac$$
and $R_0 $ satisfies \ImTzz\ with $R$ replaced by $R_0$ and $S$ by
$S_0$.

The equation \hamjac\ is just the classical Hamilton-Jacobi equation
and is
equivalent to the classical equations of motion once we introduce the
classical
trajectory through the equations,
$${1\over 16\pi}\dot Z_{\pm}(\tau ,\s ) = \Pi_{ \pm}= {\d S_{
-1}\over\d \z_-
}\biggr{|}_{\z_\pm =Z_\pm\atop f_i=F_i}\qquad \qquad{1\over 4\pi}\dot
F_i(\tau,\s)=\Pi_f= {\d S_{ -1}\over\d f_i }\biggr{|}_{\z_\pm
=Z_\pm\atop
f_i=F_i}.\eqn\ftraj$$

Differentiating these equations with respect to time, and the spatial
integral
of
\hamjac\ with respect to $\z_\pm$ and $f_i$, we obtain the classical
equations
of
motion for the Liouville-like theory: \foot{I.e. the classical CGHS
theory
modified by the measure corrections that are of order $\k\hbar$ and
have in our
large N theory been promoted to the status of classical effects.}
$$\ddot Z_--Z''-8e^{Z_+}=0,\qquad \ddot Z_+-Z_+''=0,\qquad \ddot
F-F''=0.$$
Substituting  \ftraj \ into \hamjac, and similarly deriving the
corresponding
equation for the $T_{01}$ constraint, one gets the classical
constraint
equations
$$\eqalign{0 =& {1\over 8} \left ( \dot Z_+ \dot Z_- + Z'_+ Z'_- +
2\left
(Z''_+ - Z''_-\right) - 8 \l^2 e^{Z_+}\right ) + {1\over 4}
\sum_{i=1}^N\left
({F'_i}^2 + {\dot F_i}^2\right ) ,\cr0 = & {1\over 2} \left [
\left(Z'_- + 2
{\partial\over \partial \s}\right) \dot Z_+ + \left(Z'_+ - 2
{\partial\over
\partial \s}\right) \dot Z_- \right ] + \sum_{i=1}^N F'_i \dot F_i
.}$$

These equations are precisely those which have been extensively
discussed
recently in
connection with Hawking radiation. {\it In our view these have to be
viewed not
as equations for  expectation values of field operators but as the
equations
for WKB trajectories.}  A  notion of time evolution emerges only
along a
trajectory, and it should be stressed that  at the level of the WKB
wave
function itself there is no Schr\"odinger time evolution since it is
still just
an (approximate) solution to the Wheeler-DeWitt equation.  This
should be
contrasted with what emerges from the Born-Oppenheimer approximation
to the
WdW equation. In (four-dimensional) minisuperspace for instance this
approximation corresponds to holding $\hbar $ fixed and doing a large
$M_P $
(i.e. compared to  matter energy density) expansion. In this case (as
first
shown by Rubakov  and Lapchinsky [\banks ]) one obtains a
Schr\"odinger
equation for the matter wave function with time evolution determined
by a
classical trajectory for the geometry (scale factor). In the 2D
dilaton-gravity
case the corresponding approximation consists of taking the large N
(or $\k$)
limit while holding $\hbar$ fixed. Again one obtains [\mc ] a
Schr\"odinger
equation for the matter wave function in a classical geometrical
background.
However it should be stressed that in this approximation there is no
back
reaction. In fact, as explained in detail in [\mc ], there is no
systematic way
of getting a Schr\"odinger equation for matter in a geometrical
background that
is backreacting to the evolution of the matter.

By contrast, in the semiclassical WKB approach  there is both time
evolution
and backreaction of the geometry to matter, but in a purely (semi-)
classical
manner,  i.e. along a WKB trajectory. Here there is no Schr\"odinger
evolution
of the matter wave function. In order to get time evolution {\it and
}
backreaction one has to pick a (classical) trajectory in the WKB
sense. As far
as we can see this is the only interpretation (starting from an exact
formulation of the corresponding quantum gravity) that is available
for the
semiclassical pictures of blackhole formation and evaporation that
have been
discussed in connection with CGHS models.

Finally let us consider the possibility of introducing time ``as
measured by"
clocks at infinity. Since we are dealing with a asymptotically flat
space time
classically we can define a boundary Hamiltonian, i.e. the ADM
energy. The
question at hand is whether there is a quantum analogue of this which
enables
us to define Schr\"odinger evolution and avoid the problems discussed
above.
The Hamiltonian of the theory may be written (going back to the $X,
Y$ field
basis) as  (slightly changing the normalization from that used
before)
$$H_0={1\over 2}\int d\s \left [ 2( \Pi_X^2-\Pi_Y^2) + {1\over
2}(X'^2-Y'^2) +
2 Y''-4\l^2 e^{(X+Y)}\right ] +H^f\eqn\ham$$
As Regge and Teitelboim[\rt] have pointed out, in order to get
Hamilton's
equations from this and therefore to interpret this object as the
time
translation generator, it is necessary to add a boundary term
$H_{\pa} $ to
compensate for the boundary contribution that arises in the course of
deriving
the equations of motion. The total classical Hamiltonian (time
translation
generator) is thus $H=H_0+H_{\pa} \approx H_{\pa}$. Let us now
consider \ham\
as an operator  Hamiltonian in the quantum theory (X, Y, f, are now
operators)
and look at  the derivation of the Heisenberg equations. Using the
commutation
relations
$[X(\s ),\Pi_X(\s ')]=i\d ( \s -\s ') $ etc. we get
$i\dot{X}=2i\Pi_X$ and
$$\eqalign{i\dot{\Pi}_X&=[\Pi_X(\s ), H_0] \cr
&={1\over 2}\int d\s '\left [X'(\s )(-i\pa_{\s}\d (\s-\s ')-4\l^2(-i
\d (\s-\s
')e^{(X+Y)(\s ')}\right ]\cr
&={ i\over 2}\left (X''(\s )+4\l^2e^{ (X+Y)(\s )}\right
)\cr}\eqn\heisenberg$$
Combining these two equations we get the equation of motion for $X$,
and
similarly for $Y$. The point of the above elementary exercise is to
highlight
the fact that in the quantum theory, unlike in the classical theory,
there is
no need for a boundary term since we are dealing with distributions.
Alternatively one may define the theory rigorously as a CFT  in a box
(see for
instance [\dijkvv]) and then take the size of the box to infinity.
Again the
irrelevance of the boundary term is apparent.

Even though such a term is not required by the Regge-Teitelboim
argument, one
may still add an operator valued surface term to the Hamiltonian and
then argue
that there is a Schr\"odinger time evolution with respect to that. In
other
words one may write,
$$i\hbar {\pa \Psi\over\pa t}=(H_0+H_{\pa})\Psi=H_{\pa}\Psi$$
(for physical states $\Psi$).  However,  because  {\it local}
operators  will
commute with a boundary Hamiltonian defined at spatial infinity,
their
expectation values in physical states will be independent of the time
defined
above.

What lessons can we draw from the discussion in this paper for the
question of
information loss?  It is our contention that this  can be posed only
if one can
derive a picture of Schr\"odinger evolution in a background geometry
that
behaves classically and backreacts to the matter evolution. But this
is
precisely what is not available to us.  As far as we can see there
are  but two
options. One is to use the Born-Oppenheimer approximation. In this
case one has
a notion of quantum mechanical states for the matter sector that
evolve
according to Schr\"odinger's equation. The background however is a
solution of
the vacuum Einstein equation and does not backreact.  On the other
hand one can
take the WKB semiclassical approximation. Here back reaction is
included and
the well-known Penrose diagram of black hole formation and
evaporation can be
justified. However both the geometry and the matter have to be
treated as  the
classical trajectories that one obtains in the WKB approximation.
There is  no
notion of Schr\"odinger evolution of the matter wave function.
Within this
context the  question of loss of quantum mechanical information
cannot even be
posed.

{\bf Acknowledgements:} SdeA would like to acknowledge the
hospitality of the
Newton Institute, Cambridge,  where some of this work was done.  He
would also
like to thank K. Kuchar, S. Mathur, J. Preskill, K. Stelle, A.
Strominger, and
L. Susskind for discussions.   In addition he would like to
acknowledge the
award of a Japan Society for the Promotion of Science fellowship and
the
hospitality of the University of Osaka.  This work is partially
supported by
the Department of Energy contract No.
DE-FG02-91-ER-40672.

\vfill\eject

\refout

\end